\documentclass[aps,prl,showkeys,tightenlines,nofootinbib, twocolumn]{revtex4-1}
\usepackage{amsmath,amssymb, amsfonts}
\usepackage{graphicx}

\newcommand{\ii}{\mathrm{i}}

\begin{document}

\title{Spontaneously broken conformal invariance in observables} 
 
\author{Rutger H.  Boels}
\email{Rutger.Boels@desy.de}
\author{Wadim Wormsbecher}
\email{Wadim.Wormsbecher@desy.de}
\affiliation{II. Institut f\"ur Theoretische Physik,\\ Universit\"at Hamburg,\\  Luruper Chaussee 149,\\ D-22761 Hamburg, Germany}

\date{\today}

\begin{abstract}
\noindent Conformal invariance is spontaneously broken in many physical systems leading to the appearance of a single massless Goldstone mode in the spectrum, the dilaton. The dilaton soft limit is shown to generically encode the action of both the dilatation and the special conformal transformation on observables. For massive on-shell legs these take the form of sub-leading soft theorems. At loop level, we show how anomalous dimensions for coupling constants and fields can be included. We illustrate the general analysis with a variety of formal and phenomenological applications.  
\end{abstract}

%\preprint{?}
% 11.25.-w 	Strings and branes
% 11.25.Db	Properties of perturbation theory 

%\pacs{11.25.Db}


% note 600 character limit for abstract! (includes spaces!)

\maketitle

\section{Introduction}
Symmetry is a central theme of modern physics. Frequently the symmetry is spontaneously broken: the ground state of the theory does not respect the symmetries of the equations of motion. Examples of theories with spontaneously broken symmetries abound, let us mention particle physics (e.g. \cite{Coleman}), condensed matter physics (e.g. \cite{Nicolis:2015sra}) and cosmology (e.g. \cite{Cheung:2007st}). 

In a relativistic theory the breaking of a continuous symmetry will lead to the appearance of one or more massless excitations in the spectrum, the Goldstone modes \cite{Goldstone:1961eq}. Basically, breaking a continuous symmetry requires a non-vanishing vacuum expectation value (vev) for one or more fields. The Goldstone modes are long-wavelength fluctuations of this vev. In the limit in which the momenta of Goldstone particles become very small, the so-called soft limit, one basically studies a symmetry variation. Goldstone particles are massless as they do not have to overcome an energy barrier. It is known (e.g. \cite{Low:2001bw} and references therein) that the Goldstone mechanism is fundamentally different for space-time symmetries which act on the space variables $x^{\mu}$ compared to the much better studied case where the symmetry only acts on fields.  

For broken internal symmetries such as the approximate chiral symmetry of quarks the soft limit of the Goldstone mode in scattering amplitudes generically vanishes. This can be used for instance to prove neutrinos are not Goldstone particles of broken supersymmetry \cite{deWit:1975th}, see e.g. \cite{Cheung:2014dqa} for a recent modern application. For spontaneously broken conformal symmetry it is shown below that the soft limit is qualitatively different. Conformal symmetries are space-time transformations which leave angles between vectors invariant; scale symmetry is a part of this. They are the most general bosonic global space-time symmetries which still allow interesting physics \cite{Coleman:1967ad}. Conformal symmetry has wide phenomenological and formal applications. The Goldstone particle of broken conformal invariance is called the dilaton. The main technical result of this letter is to show how for general observables the dilatation operator and the special conformal transformation appear in the dilaton soft limit. Where we are aware of related results in the literature, this will be mentioned below. 

\section{Dilaton soft limit from the Ward identity}

Particles are represented by quantum fields with a mass and a spin. Here they will also have a definite scaling dimension, i.e. under $x  \rightarrow e^{\lambda} x $ they will transform as
\begin{equation}\label{eq:defscalingdim}
\mathcal{O}_i(x) \rightarrow e^{d_i \lambda}\mathcal{O}_i\left(e^{\lambda} x \right)
\end{equation}
where $d_i$ is the scaling dimension. Under special conformal transformations,
\begin{equation}
x^{\mu} \rightarrow  x^{\mu} + 2 (a \cdot x) x^{\mu} - a^{\mu} x^2  + \mathcal{O}(a^2)
\end{equation}
the operators will then transform as (e.g. \cite{DiFrancesco:1997n}), 
\begin{equation}\label{eq:specconftrafoinf}
\mathcal{O}_i(x) \rightarrow \left( 1 + 2 (a \cdot x) d_i - a^{\mu} x^{\nu} S_{\mu\nu}  \right) \mathcal{O}_i (x') +  \mathcal{O}(a^2)
\end{equation}
where $S_{\mu\nu}$ is the Lorentz generator of the representation of the field $\mathcal{O}$. Assume there is a scalar field in the theory, denoted $\xi$, which acquires a vev,
\begin{equation}
\langle \xi \rangle = a \neq 0
\end{equation}
and that the theory has a stable vacuum at this point. This vev breaks conformal symmetry if the scaling dimension of $\xi$ is not zero. The excitations of the field $\xi$, denoted $\tilde{\xi}$,
\begin{equation}
\xi = a + \tilde{\xi}
\end{equation}
are the Goldstone modes and are massless. The action of the theory is invariant under the dilatation of the original field, which can be written as
\begin{equation}\label{eq:symfordilaton}
\tilde{\xi}  \rightarrow \tilde{\xi} + \lambda\, a +  \lambda  \, \tilde{\xi}  + \lambda\, x \partial_x \, \tilde{\xi} + \mathcal{O}(\lambda^2) 
\end{equation}
The Ward identity of this transformation can be derived in the standard fashion from the path integral, apart from the fact that the source term for the quantum field $\tilde{\xi}$
\begin{equation}
\sim \ii \int d^D x \, J \tilde{\xi}
\end{equation}
is not invariant under the symmetry \eqref{eq:symfordilaton}. Instead, this inserts an amputated zero momentum dilaton field. The Ward identity reads
\begin{equation}\label{eq:schematicward}
\sum_{i=1}^{n} \Delta_i G_{n}(p_1, \ldots, p_n) = a \ii \tilde{F}_n(p_1, \ldots, p_n)
\end{equation}
where the left hand side is the action of the dilatation operator on the connected correlator of $n$ fields and on the right is a correlator with the same field content plus an additional soft dilaton, times the vev. This soft dilaton identity is exact modulo anomalous dimensions for fields and coupling constants, see below. This result is implicit in \cite{Coleman} and the derivation resembles that of the Callan-Symanzik equations \cite{Callan:1970yg}\cite{Symanzik:1970rt}. 

Equivalent forms for this identity can be derived. For a connected, amputated, 1PI correlator with massless fields of scaling dimension $d$ at tree level in a theory with only one scale $m$ one obtains for instance,
\begin{multline}\label{eq:samplewardid}
\left[-  m \frac{\partial}{\partial m}  + n \left(d - \frac{(D -2)}{2} \right) \right] G^{\textrm{amp}}_{n}(p_1, \ldots, p_n) = \\  a  \ii\, F^{\textrm{amp}}_n(p_1, \ldots, p_n)
\end{multline}
At loop level and with massive on-shell fields additional care is needed, see below. The derivation of the Ward identity for dilatations can be repeated for special conformal transformations. Pushing through the derivation and using $S_{\mu\nu}=0$ for scalar fields gives
\begin{equation}
\sum_{i=1}^{n} K^{\mu}_i G_{n}(p_1, \ldots, p_n) = a \ii \, \frac{\partial}{\partial p^{\tilde{\xi}}_{\mu}} \tilde{F}_n(    p_1, \ldots, p_n)
\end{equation}
where on the left-hand side one has the action of the special conformal transformation, while on the right hand side one must insert a dilaton, amputate, take its momentum derivative and then take its momentum to zero. The analog of the Callan-Symanzik equation in this case can be found in \cite{Parisi:1972zy}.

\subsection{Extension to massive matter}
The subtlety in the derivation of equation \eqref{eq:samplewardid} for massive legs is in the amputation step. The inverse propagator is generically not covariant under dilatations but proportional to a soft dilaton insertion itself,
\begin{equation}
\Delta_i G_{1PI}^{(2)} =  a \ii \, F_{1PI}^{(2)}
\end{equation}
For massless matter the right hand side is always proportional to $G$ itself, but it generically is not for massive on-shell matter. One can see this directly already for tree level amplitudes with massive matter in a spontaneously broken theory: there is a soft divergence from Feynman graphs where the dilaton connects to the outside massive leg directly. Note that the mass of the matter also enters into the scattering amplitude by the on-shell condition $p^2=m^2$. These two problems can be made to cancel each other, leaving a version of the result in equation \eqref{eq:samplewardid} valid for scattering amplitudes. 

To see this, consider the Feynman graph expansion of an observable with fields of canonical scaling dimension and massive matter. Take the momentum of the dilaton to be $\lambda p_s$ The divergent graphs contributes
\begin{equation}\nonumber
A_{n+1}(p_s, X)|_{\textrm{div.}} = \sum_{i \in \textrm{massive}} J_3 \frac{1}{2 \lambda p_s \cdot p_i} J_n(p_i + \lambda p_s, X)
\end{equation} 
For $(p_i + \lambda p_s)^2 = m^2$ the current $J_n$ on the right hand side becomes the $A_n(X)$ scattering amplitude by unitarity. The (twice amputated) current $J_3$ in the limit $\lambda \rightarrow 0$ is 
\begin{equation}
\lim_{\lambda \rightarrow 0} J_3 = \frac{1}{a} m \frac{\partial}{\partial m} J_2 = \frac{1}{a} 2  m_i 
\end{equation} 
where the last equality holds at tree level. At first order in $\lambda$ this current contains a part with derivative of delta function support. Moreover, the explicit momentum flowing through a given propagator depends on momentum conservation. First take the momentum running through a given propagator to be the sum of momenta on the side of the propagator which contains an arbitrarily chosen particle $1$.  This fixes momentum routing for each propagator. The $\lambda$ dependence of a given propagator now depends on wether the soft dilaton appears on the same or opposite side of the graph compared to particle $1$. Any other momentum routing will lead to an expression which vanishes up to derivative of delta function support. 

The Laurent series in $\lambda$ for the scattering amplitude can now be studied, adapting an argument from \cite{Bern:2014vva}. The leading divergent part is fixed by unitarity. For the subleading, $\lambda^0$, part, we obtain
\begin{equation}\label{eq:subleading}
A_n(\lambda p_s, X)|_{\lambda^0} = \frac{m}{a} \frac{\partial}{\partial m} A_n(X) 
\end{equation} 
up to terms which have derivative of delta function support: this is a distributional equation. All massive momenta must be expressed as
\begin{equation}\label{eq:massmom}
p_i = p^{\flat}_i + \frac{m_i^2}{2 p \cdot p_i} p_s
\end{equation}
Note $(p^{\flat}_i)^2=0$ and this relation solves $p_i^2 = m_i^2$ explicitly. The mass derivative can hit internal propagators, where its effect corresponds on the left hand side to the expansion of the momenta in the propagators. If the soft dilaton momentum is constrained to satisfy \cite{Kiermaier:2011cr}
\begin{equation}\label{eq:kiermaierconstraint}
\sum_{i} \frac{m_i }{2 p_s p_i} = 0
\end{equation} 
then both the leading divergence cancels and the mass dependence drops out of the delta function. We conjecture that the sub-sub-leading soft dilaton limit in the case of massive on-shell matter gives the special conformal transformation. 

\subsection{First applications}

Consider the Higgs effect. The scalar vev which breaks gauge symmetry could also break a scale symmetry. The dilaton is then necessarily uncharged under the residual gauge group: it must be a massless singlet. Its low energy couplings are dominated by the soft limit derived above. Hence the dilaton-quark couplings are proportional to the induced masses of the matter in the theory. In the low energy limit, this observation combine with Lorentz symmetry completely determines the couplings to be proportional to the Yukawa couplings. Moreover, particle decay into two dilatons is prohibited: the three point decay amplitude for instance would be proportional to the massive particle mass, whose soft limit does not vanish. As an illustration consider the standard model, with a Higgs potential modified to
\begin{equation}
V = \frac{\lambda}{2} (\vec{H}^2 - b \xi^2)^2
\end{equation}
where $\xi$ is a singlet under all standard model gauge groups. This is a special limit of the general idea of a Higgs portal \cite{Patt:2006fw} (see also \cite{Meissner:2006zh}). The requirement of a flat direction fixes one constraint for the coupling constants among the $3$ scaling invariant terms one can make out of two scalar fields: up to a field redefinition this potential is the most general. This potential has a degenerate vacuum, 
\begin{equation}
\langle \vec{H}^2 \rangle = (b a)^2 \qquad \langle \xi \rangle = a
\end{equation}
Expanding around the vacuum in standard unitary gauge and using the field redefinition
\begin{equation}
h(x) \rightarrow  \frac{(h'(x) + b \xi')}{\sqrt{1 + b^2}} \quad 
\xi \rightarrow \frac{(\xi' - b h'(x) ) }{\sqrt{1 + b^2}}
\end{equation} 
gives canonical kinetic terms for $h'$ and $\xi'$ and a potential
\begin{equation}\label{eq:dilhiggs}
V(h', \xi') = \frac{1}{2} (h')^2 \left(m + \lambda (b \xi' + (1-b^2) h')  \right)^2
\end{equation}
This shows $\xi'$ is the massless dilaton and $h'$ is the massive 'Higgs', with mass $m= 2 a b \sqrt{1+b^2} \lambda$. The field shift induces dilaton-matter couplings, aligned along the Yukawa interactions, as described. For $b=1$ this is exactly Coleman's example  \cite{Coleman} to illustrate spontaneous scale symmetry breakdown. It is instructive to verify the general statements for Green's functions and amplitudes above when $b=1$. See \cite{Robens:2015gla} for a discussion of collider constraints on this theory. 

As a second application, consider the dilaton soft limit in tree level string theory. String theory amplitudes are governed by the closed string coupling constant and the string scale $\alpha'$. The dilaton soft limit was discussed in the single brane case in \cite{Ademollo:1975pf}. Comparing to the general result above, we immediately identify the string theory dilaton as the Goldstone of spontaneous scale symmetry breakdown. Moreover, the natural scaling dimension of the closed modes is zero, while that of the open modes is $\frac{(D-2)}{4}$. In the field theory limit our general result is fully consistent with the novel subleading soft graviton theorems of \cite{Strominger:2013jfa} \cite{Bern:2014vva} (see also \cite{DiVecchia:2015oba}). 

As a third application, consider a D-brane system of stacks of parallel branes, separated along an orthogonal direction which breaks scale and gauge symmetry. In the field theory limit this is the so-called Coulomb branch of the effective field theories living on the world-volume of these branes. There is a precise map \cite{Boels:2010mj} from this setup to higher dimensional Yang-Mills theories without symmetry breaking. Tracing the recently discovered sub-leading soft gluon theorems \cite{Casali:2014xpa} \cite{Bern:2014vva} through this map shows equivalence to \eqref{eq:subleading} for the dilaton. Equation \eqref{eq:massmom} is essential here. 

Note that the results above are closely related to consistency conditions in cosmology, see \cite{Maldacena:2002vr} \cite{Hinterbichler:2013dpa}  \cite{Kundu:2015xta}. There, for single field inflation, the role of the dilaton is played by the inflaton and the broken scale symmetry translates to time translation invariance. It would be highly interesting to explore this connection further. 

\subsection{Passing to the quantum level}
Quantum effects can break conformal invariance. The question relevant for this letter is if this breaking is explicit or (can be upgraded to be) spontaneous. It is easy to guess what the Ward identity would be if the latter scenario is realised: the dilatation operator is extended to include the anomalous dimensions of fields and coupling constants, see \cite{Tamarit:2013vda} for a one loop discussion. 

There is a dimensional regularisation scheme \cite{Englert:1976e}, which preserves conformal invariance. For the model listed in equation \eqref{eq:dilhiggs} this reads in the case $b=0$,
\begin{equation}\label{eq:scalinvtheorynaive}
\mathcal{L} = \frac{1}{2} \partial_{\mu} \xi \partial^{\mu} \xi +  \frac{1}{2} \partial_{\mu} \phi \partial^{\mu} \phi - \frac{1}{2} \left(\frac{\lambda \xi}{a} \right)^{\frac{4}{D-2}} \phi^2 a^{\frac{4}{D-2}}
\end{equation}
The exponential is well-defined in non-integer dimensions as long as the theory is expanded around a nontrivial $\xi$ vev, $\xi = a + \tilde{\xi}$. This leads to an infinite series of interaction terms and the $\phi$ field acquires a mass
\begin{equation}
m_{\phi} = (\lambda a)^{\frac{2}{D-2}} 
\end{equation} 
which is kept fixed in perturbation theory. This theory is conformal invariant, with scaling dimension $\frac{D-2}{2}$ for the fields and $0$ for the coupling constant. The Lagrangian contains an infinite series of, generically, not power-counting renormalisable terms which are evanescent. Explicit computation, \cite{Shaposhnikov:2009nk} \cite{Gretsch:2013ooa}, seems to point to non-renormalisability at higher loop order. 

Even at one loop the theory above contains what will be called subdivergences: there are UV divergent graphs with an arbitrary number of external $\xi$ legs, whose overall value is $\frac{1}{\epsilon} \epsilon \sim 1$. These finite terms feed into the observed higher loop nonrenormalisability. In the standard regularisation these subdivergences are absent. We propose that they can be systematically cancelled by using anomalous dimensions as additional renormalisation constants in the above scheme. Note that this decouples the direct relation between the $\frac{1}{\epsilon}$ divergence and the renormalisation logs. These features all are closely related to two loop divergences in general relativity \cite{Bern:2015xsa}. 

The key is that the dilaton soft limit of the above theory, potentially including a renormalisation scale $M$,
\begin{equation}\label{eq:noanomaly}
\lim_{p_s \rightarrow 0} F(p_s, X) = \frac{1}{a} \left( - m \frac{\partial}{\partial m} - M \frac{\partial}{\partial M}\right) G(X)
\end{equation}
is an identity on the level of Feynman graphs: anomalous dimensions can not arise. The divergences can always be cancelled by scale invariant counterterms which have an exactly dimensionless coupling constant. The Ward identity derivation above guarantees it is annihilated by the operator on the right of equation \eqref{eq:noanomaly}. 

The Ward identity for scale transformations can include anomalous conformal dimensions for coupling constants and fields. In the regularisation used here, these dimensions enter directly into the action, e.g.
\begin{multline}\nonumber
\xi_0 \rightarrow a \left(\frac{\xi_r}{a}\right)^{\frac{D-2}{D-2 - 2 \delta_{\xi}}} \qquad \lambda_0 \rightarrow \lambda_r \left(\frac{\xi_r}{a}\right)^{-\delta_{\lambda} (\frac{D-2}{2} - \delta_{\xi})^{-1}}
\end{multline}
In addition, one introduces re-scalings of coupling constants and fields. Combined with pure $\xi$ and $\phi$ type counterterms with coupling constants $\lambda_a$ and $\lambda_k$, their anomalous dimensions, $\delta_{\lambda_a}$ and $\delta_{\lambda_k}$ and rescalings this leads to a total of $10$ counterterm parameters around $4$ dimensions. The soft limit now yields
\begin{equation}\label{eq:newanomaly}
\propto \left(m \frac{\partial}{\partial m} + \sum_{i=1}^3 \delta_{\lambda_i} \lambda_i  \frac{\partial}{\partial \lambda_i} + n_{\phi} \delta_{\phi} + n_{\xi} \delta_{\chi}\right) G(X_r)
\end{equation}
The field and coupling constant rescalings will cancel the $\frac{1}{\epsilon}$ poles, while the \emph{finite} anomalous dimensions cancel subdivergences. The $5$ finite parts of the field and coupling constant rescalings are then fixed by renormalisation conditions. Derivatives of coupling constants set to zero (e.g. $\lambda_a$) insert corresponding operators.

As an example of non-trivial $\beta$ functions, consider one loop in $D=4-2\epsilon$ dimensions where the divergent integrals all have a tadpole or bubble topology. The field renormalisation constants and their anomalous dimensions are not needed to cancel any divergences. Setting  
\begin{equation}\nonumber
\delta_{\lambda_a} = \frac{ 12 \lambda^4}{(4 \pi)^{2} \, \lambda_a}  \qquad \delta_{\lambda} = \frac{\lambda_k + 4 \lambda^2}{(4 \pi)^{2}}  \qquad  \delta_{\lambda_k} =  \frac{12 \lambda^4+  3 \lambda^2_k}{\lambda_k (4 \pi)^{2}} 
\end{equation}
cancels the subdivergences. This reproduces the beta functions computed in standard four dimensional regularisation without symmetry breaking. 

As an example of non-trivial anomalous dimensions, consider one loop in $D=6-2\epsilon$ dimensions and focus on the only momentum dependent divergences in the quadratically subdivergent bubble graphs with many external legs. All subdivergences in this class can be cancelled by anomalous dimensions of $\phi$ and $\xi$ as
\begin{equation}
\delta_{\phi} =  2 \delta_{\xi} =  \frac{ \lambda^2}{6 (4 \pi)^{3}}
\end{equation}
These anomalous dimensions are the same as those in the standard regularisation without symmetry breaking. In addition, there are subdivergent graphs for the coupling constants of $\xi^{\frac{2 D}{D-2}}$ and $\phi^2 \xi^{\frac{4}{D-2}}$ which lead to $\beta$ functions. There are however also subdivergent triangle graphs with four external $\phi$ fields which cannot be cancelled in this fashion; the non-standard propagator correction vertices could cancel the generated divergences at higher loops. 

We conjecture, but do not prove, that with anomalous dimensions included the regularisation scheme above defines a consistent, renormalisable quantum field theory with spontaneously broken scaling symmetry whose observables coincide with traditional results up to evanescent terms. 

\subsection{Discussion}
We have shown how the dilaton soft limit encodes the action of the conformal generators. Several immediate physical applications have been highlighted in the text. It would be highly interesting to push this program further for instance in the context of the AdS/CFT or dS/CFT correspondences, where the latter has cosmological applications. Exploring renormalisable quantum field theories with (spontaneously) broken conformal invariance at loop level may have very wide formal and phenomenological applications. This will shed new light on the role of scale invariance in physical systems. This article presents a first step in exploring the consequences of the general class of broken space-time symmetries for observables. Beyond local conformal symmetry this class further includes diffeomorphism and Poincare invariance, where the latter is broken in general membrane systems such as low energy string models for mesons \cite{Aharony:2013ipa} or D-brane stacks.

\section*{Acknowledgements}
We would like to thank Arthur Lipstein and Alexander Westphal for discussions. RB would like to thank Humboldt and Nottingham universities for the opportunity to present the results reported here. This work has been supported by the German Science Foundation (DFG) within the Collaborative Research Center 676 ``Particles, Strings and the Early Universe".

\bibliographystyle{apsrev4-1}

\bibliography{dilaton.bib}

\end{document}